\documentclass[aps,physrev,preprint,groupedaddress]{revtex4-2}

\usepackage{url}
\usepackage{natbib}
\usepackage{amsfonts}
\usepackage{amsmath}
\usepackage{amssymb}
\usepackage{braket}
\usepackage{nicefrac}
\usepackage{array}
\usepackage{booktabs}
\usepackage{tikz}
\usepackage{quantikz}
\usepackage{rotating}
\begin{document}

\title{\textbf{Quantum Computing Beyond Ground State Electronic Structure: A Review of Progress Toward Quantum Chemistry Out of the Ground State} 
}% 

\author{Alan Bidart$^{1}$}
\thanks{These authors contributed equally to this work.}
\author{Prateek Vaish$^{1}$}
\thanks{These authors contributed equally to this work.}
\author{Tilas Kabengele$^1$}
\author{Yaoqi Pang$^1$} 
\author{Yuan Liu$^{2,3,4}$} 
\author{Brenda M. Rubenstein$^{1,5,6}$}
\email{brenda\_rubenstein@brown.edu}
\affiliation{$^1$Department of Chemistry, Brown University, Providence, RI, USA, 02912}
\affiliation{$^2$Department of Electrical and Computer Engineering, North Carolina State University, Raleigh, NC USA, 27606}
\affiliation{$^3$Department of Computer Science, North Carolina State University, Raleigh, NC, USA, 27606}
\affiliation{$^4$Department of Physics, North Carolina State University, Raleigh, NC, USA, 27607}
\affiliation{$^5$Department of Physics, Brown University, Providence, RI, USA, 02912}
\affiliation{$^6$Data Science Institute, Brown University, Providence, RI, USA, 02912}

\date{\today}

\begin{abstract}
Quantum computing offers the promise of revolutionizing quantum chemistry by enabling the solution of chemical problems for substantially less computational cost. While most demonstrations of quantum computation to date have focused on resolving the energies of the electronic ground states of small molecules, the field of quantum chemistry is far broader than ground state chemistry; equally important to practicing chemists are chemical reaction dynamics and reaction mechanism prediction. Here, we review progress toward and the potential of quantum computation for understanding quantum chemistry beyond the ground state, including for reaction mechanisms, reaction dynamics, and finite temperature quantum chemistry. We discuss algorithmic and other considerations these applications share, as well as differences that make them unique. We also highlight the potential speedups these applications may realize and challenges they may face. We hope that this discussion stimulates further research into how quantum computation may better inform experimental chemistry in the future.  
\end{abstract}

\maketitle

\section{INTRODUCTION}

The field of quantum chemistry, which leverages the principles of quantum mechanics to solve chemical problems~\cite{szabo1996modern},
has emerged over the past few decades as one of the most vital areas of modern science because of its unparalleled ability to grant deep insights into the chemical processes that govern life~\cite{friesner2005ab,kirchner2007theoretical}, catalysis~\cite{norskov2011density}, materials~\cite{marzari2021electronic}, and many other natural phenomena. Quantum chemistry has shed light on the atomistic details of such important phenomena as photosynthesis, superconductivity, and the central dogma, enabling researchers to not only understand, but to control many phenomena. While many conceive of quantum chemistry as focused on solving the time-independent Schr\"{o}dinger Equation for electronic ground states, chemistry - and by extension, quantum chemistry - is concerned with a much wider and richer variety of phenomena often directly observed in the lab including vibrational and rotational motion, chemical reaction mechanisms, reactive dynamics, and kinetics. The modeling of such phenomena on classical hardware is often predicated on the full or partial solution of the time-dependent Schr\"{o}dinger Equation, but a wide variety of approximations to this full solution have arisen (see Section \ref{quantumchemnoground}). From this perspective, ground state electronic structure is like the foundation for a more elaborate mansion: it is a necessary piece upon which more visible, enchanting, and in this case, experimentally meaningful, accoutrements can be built.

Quantum computation holds the promise of impacting - and potentially transforming - not just the foundation, but the full edifice of quantum chemistry. By leveraging quantum systems such as superconducting qubits~\cite{kjaergaard2020superconducting} and trapped ions~\cite{bruzewicz2019trapped} with such essential quantum properties for computation as superposition and entanglement to model other quantum systems, quantum computers have the potential to polynomially, if not exponentially, accelerate the solution of quantum chemical problems~\cite{feynman1982simulating,lee2022there}. Determining the exact quantum ground state of a system on classical hardware, typically scales exponentially with system size, as it involves manipulating an exponentially growing number of quantum states. However, because of the properties of superposition, one can encode an exponential number of states in a linear number of qubits on a quantum computer. Leveraging entanglement, quantum computers can moreover perform complex operations on many qubits at once, dramatically reducing the operational cost of different computations~\cite{nielsen2010quantum}. While exciting classical advances have been made that enable many ground state electronic structure problems (such as main group quantum chemistry) to effectively, although not exactly, be solved with polynomial cost, the promise that quantum computers could solve these same problems at lower cost and other problems (such as strongly-correlated, multimetallic enzymes that cannot be readily solved by current classical algorithms) at polynomial cost motivates the field~\cite{bauer2020quantum,cao2019quantum,Mcardle_2020}. Currently, a variety of quantum computing techniques, including Variational Quantum Algorithms (VQA)~\cite{peruzzo_variational_2014,tilly2022variational}, quantum Krylov methods~\cite{yoshioka2025krylov,cortes_quantum_2022}, and quantum Monte Carlo methods~\cite{huggins2022unbiasing} have studied the ground states of small molecules, including water, molecular dimers, and hydrogen chains. These demonstrations show that even modern Noisy Intermediate-Scale Quantum (NISQ) devices~\cite{preskill2018quantum} without error correction can solve quantum chemistry problems, and with further, likely, advances in error correction, measurement techniques, wave function initialization, and quantum ansatzes, can hold great promise.

Nonetheless, despite the scientific and practical importance of quantum chemistry simulations beyond ground state modeling, far fewer demonstrations of more general quantum chemistry applications have been performed to date. We view these ground state demonstrations as building the foundation for quantum computation in quantum chemistry, yet leaving it bare, without the edifice and furnishings that inspire many chemists.  Some of this state-of-affairs owes to algorithmic and hardware challenges, but we take the position that, with community effort and focus, many of these hurdles can be surmounted. This is especially important since research suggests that the greatest speedups for quantum chemistry problems may apply to quantum dynamics~\cite{lee2022there,kassal2008polynomial,stroeks2024solving}. 

In this review, we therefore focus on how quantum computation is impacting the field of quantum chemistry in the broadest sense: while the majority of reviews have focused on ground state electronic structure, this review will focus on developments in quantum computation related to other aspects of quantum chemistry, including reaction mechanisms, Born-Oppenheimer molecular dynamics, quantum dynamics, and finite temperature electronic structure. In so doing, we aim to highlight the potential impact quantum computing may have on practical experimental chemistry. We furthermore aim to underscore the commonalities and differences that emerge among the quantum algorithms designed to solve different genres of quantum chemical problems, and in particular, how these different algorithms are expected to scale relative to the quantum computational resources projected to be available in the coming years. 

\section{A PRIMER ON QUANTUM COMPUTATION}
\label{complexity}

To assess the potential for quantum computation in quantum chemistry, the first step is to establish a framework that can be used to compare classical and quantum algorithms. One of the most widely accepted metrics is computational complexity, which captures how the resources---usually space and time---needed to solve a problem scale with a problem's size.
Throughout this manuscript, we will focus on quantifying how using quantum computers to solve common genres of computational chemistry problems impacts their time-scaling.  In this section, we frame the search for quantum advantage as a question in complexity theory, explain how the circuit model of quantum computation maps onto time complexity, and review key quantum algorithms that serve as subroutines in the works surveyed.

\subsection{Quantum Complexity Theory}
Computational complexity theory provides a framework for describing which problems can be solved efficiently. 
In the classical setting, the complexity class \textbf{P} (\textbf{P}olynomial time) consists of problems solvable in polynomial time on a deterministic machine. The class \textbf{BPP} (\textbf{B}ounded-error \textbf{P}robabilistic \textbf{P}olyniomial time) extends \textbf{P} by including probabilistic algorithms that succeed with high probability.
\textbf{BPP} is often taken as a reasonable proxy for what is classically tractable. Since many problems in quantum chemistry, such as Hamiltonian simulation, are believed to fall outside \textbf{BPP}, the development of computing capabilities beyond that of classical computers is important to chemists.

Quantum computation promises to enlarge the computation landscape.
The \textbf{BQP} (\textbf{B}ounded-error \textbf{Q}uantum \textbf{P}olynomial time) class contains problems solvable efficiently with quantum algorithms, and it is widely believed that $\textbf{BPP} \subseteq \textbf{BQP}$ ~\cite{watrous2012quantum,kitaev2002classical}.
This containment relationship implies that any classically-tractable problem is also quantum-tractable.
However, depending on whether the containment is strict or not, there might be problems that only quantum computers can solve efficiently.
To put into perspective the challenge of elucidating the nature of this containment, it is worth noting that whether \textbf{P} is strictly contained in \textbf{BQP} remains a difficult open question.

We use the term ``quantum speedup" to describe a quantum algorithm that can solve a problem with a quantifiable complexity-theoretic improvement with respect to the best possible classical solution.
If the speedup is such that it yields a polynomial-time quantum solution to a problem where there exists no efficient classical solution, we say that the quantum algorithm is an instance of ``quantum advantage".
In other words, the quest for quantum advantage can be described as the search for problems at the intersection of \textbf{BQP} and the complement of \textbf{BPP}.
Answering this question involves both designing an exponentially faster algorithm and proving that there are no efficient classical-only implementations.
For example, quantum search algorithms, though proven to be asymptotically optimal, do not qualify because they only offer a quadratic speedup over their classical counterpart.
Factoring algorithms such as Shor’s famous algorithm do not qualify either because factoring has yet to be formally proven classically-intractable---although this is widely believed to be the case \cite{manenti2023quantum}.

Outside of formal complexity theory, the terms quantum advantage and quantum speedup are often used more pragmatically. 
In the next sections, we focus on describing chemically relevant problems for which quantum methods promise to provide meaningful improvements in runtime, scaling, or resource requirements, rather than on strict separations between complexity classes. 
Accordingly, we will use the term quantum advantage to refer to algorithms that offer efficient quantum solutions to problems in computational chemistry that are practically intractable for classical methods, even when those methods scale polynomially with system size.

\subsection{The Quantum Circuit Computational Model}
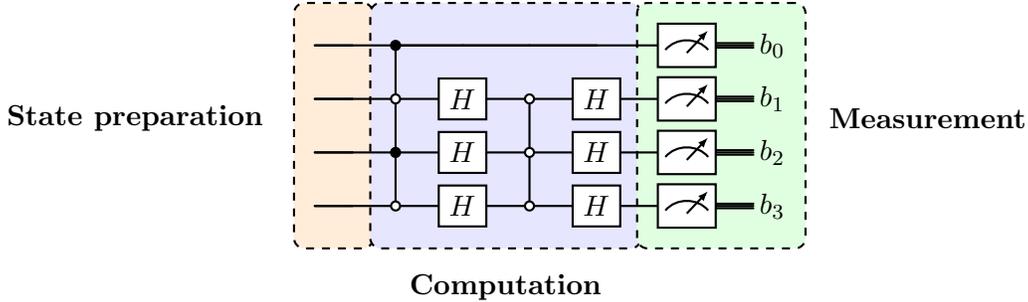
\begin{figure}
    \centering
    \begin{quantikz}[row sep={0.8cm,between origins}]
\gategroup[wires=4,steps=2,style={dashed,rounded corners,fill=orange!15},background,label style  ={anchor=east,xshift=-0.8cm, yshift=-1.75cm}]{\textbf{State preparation}}
\qw
&
\qw
&
\ctrl{0} \gategroup[wires=4,steps=4,style={dashed,rounded corners,fill=blue!10},background, label style  ={anchor=south, yshift=-4.3cm}]{\textbf{Computation}} &
\qw & \qw &
\qw
 &
\meter{} 
\gategroup[wires=4,steps=2,style={dashed,rounded corners,fill=green!12},background, label style  ={anchor=west,xshift=+1.3cm, yshift=-1.75cm}]{\textbf{Measurement}} & \cw \: b_0 \\[-2.5pt]
 &
 \qw
&
\octrl{0} &
\gate{H} & 
\octrl{2}  & \gate{H} &
\meter{} & \cw \: b_1 \\[-2.5pt]
 &
 \qw
&
\ctrl{0} &
\gate{H} & 
\octrl{0}  & \gate{H} &
\meter{} & \cw \: b_2 \\[-2.5pt]
 &
 \qw
&
\octrl{-3} &
\gate{H} & 
\octrl{0}  & \gate{H} &
\meter{} & \cw \: b_3
\end{quantikz}
    \caption{The three stages of a quantum circuit. A computational bottleneck in any of these three stages will directly compromise the performance of a quantum algorithm. For instance, if either of these stages has superpolynomial scaling with respect to system size, then the resulting algorithm will not be considered quantum-tractable. The computation step in this example features single-qubit Hadamard ($H$) gates and multi-qubit controlled gates, which combine to form the Grover iterator---a key primitive in search algorithms---associated with the $\ket{1010}$ state.
    }
    \label{fig:circuit_three}
\end{figure}

Most of the works discussed in this review make use of the circuit model to discuss quantum computation. In this model, quantum algorithms are expressed in terms of a circuit (see Figure \ref{fig:circuit_three}) in which
\begin{itemize}
    \item Qubits are represented as wires, and
    \item Quantum operators are depicted as gates.
\end{itemize}
At the start of any algorithm, an initial state is assumed to be ``prepared'' during a process known as state preparation. Despite the wires being separated in the circuit diagram, the state stored by the system of all wires is typically not a product of states, which means that wires could be entangled with each other, including in the prepared state. Then, the initial state undergoes transformations dictated by the quantum gates, read from left to right. Each gate acts on either the entire system---all of the wires---or a subsystem---a subset of wires---denoted by the shape of the gate. The final component is measurement, which can take place on a given wire or set of wires either at the end of or throughout the computation. A measurement collapses wires to one of the eigenstates of the measurement basis. If no other information is given,  the initial state for $n$ qubits is assumed to be $\ket{0}^{\otimes n}$, often abbreviated as $\ket{0}^n$, and all measurements are assumed to take place in the $Z$-basis, which collapses $m$ measured qubits to a ``computational basis state'' of the form $\ket{b_0b_1 \dots b_{m-1}}$, where $b_i \in \{0, 1\}$ for $i \in [0, m-1]$.

Other quantum computational models have been proposed and have seen industrial applications, such as measurement-based quantum computing \cite{PhysRevLett.86.5188, sakaguchi2023nonlinear} and adiabatic quantum computing \cite{farhi2001quantum, king2021scaling}. However, the circuit model is known to be universal \cite{barenco1995elementary}, which means that problems that can be solved efficiently in one of the other models can also be solved efficiently in the circuit model. 

\subsubsection{Complexity in a quantum circuit}
Assuming the state preparation stage is trivial, the time complexity of a quantum algorithm is equal to the depth of its circuit representation multiplied by the number of times the circuit must be run to reach a user-defined confidence in the result. To calculate the depth of a circuit, one often starts with a set of gates whose ``timestep cost" is set to a constant that does not vary with system size. It is standard practice to select a universal gate set, a finite collection of gates capable of efficiently approximating any quantum operation \cite{kitaev1997quantum}.
The number of timesteps required to run each non-basis gate in the circuit can then be calculated by decomposing it into the chosen basis set. After simplifying and parallelizing quantum operations as much as possible, the circuit's depth will correspond to the minimum number of ``timesteps'' needed to run the circuit. Finally, similar to a probabilistic classical algorithm, the circuit might need to be re-run multiple times to match a target confidence. 

Thus, the time complexity of the algorithm is commonly expressed using Big \textit{O} notation in terms of system size and desired accuracy.

If the state preparation required for an algorithm is nontrivial, this can significantly impact the algorithm’s overall complexity. The same applies to quantum gates that lack a straightforward or efficient decomposition into elementary operations. In these cases, we say that the algorithm depends on a \textit{black box} or \textit{oracle} to perform state preparation or computation \cite{nielsen2010quantum}. Oracles serve as placeholders for operations we currently do not know how to implement efficiently or information we expect to extract from an external system coupled to the quantum computer. 
Dependence on an oracle introduces the ``query" complexity metric, which captures how the number of times an oracle needs to be queried scales. Query complexity can be converted to time complexity only once a time complexity for the oracle itself is determined.
Additional complexity can also be introduced via the algorithm's measurement scheme. For example, as we move from Noisy Intermediate-Scale Quantum technology to fault-tolerant quantum architectures, popular error-correction and fault-tolerance schemes require a non-trivial amount of mid-circuit measurements \cite{cain2024correlated}. 

\section{Common Quantum Simulation Algorithms for Quantum Chemistry}

\subsection{Ground State Energy Estimation}
\label{ssec:groundstate}

The accurate determination of molecular ground states is the cornerstone of quantum chemistry, providing fundamental insights into chemical properties and reactivity \cite{szabo1996modern,helgaker2013molecular}. 
Classical methods like Density Functional Theory (DFT)~\cite{hohenberg_inhomogeneous_1964} and Coupled Cluster (CC) theory~\cite{cizek_correlation_1966, paldus_correlation_1972,musial2008multireference} are workhorses in this field, but their application is often limited by unfavorable scaling or limited accuracy for systems with strong electron correlation or for calculating excited electronic states. Quantum computers offer a promising path to circumventing these limitations by directly simulating the quantum nature of molecules using quantum architectures. In recent years, significant progress has been made transitioning from proof-of-principle examples to demonstrating chemical accuracy on near-term quantum devices.

The starting point for any quantum simulation of a chemical system is the electronic Hamiltonian. To represent it on a quantum computer, the Hamiltonian's fermionic annihilation ($a$) and creation ($a^{\dagger}$) operators must be mapped to qubit operators.
The Jordan-Wigner (JW) transformation is a standard choice that maps them in the following way:
\begin{align}
a_j &\mapsto \frac{1}{2} (X_j + iY_j) \bigotimes_{k<j}Z_k \nonumber \\
 a_j^{\dagger} &\mapsto \frac{1}{2} (X_j - iY_j) \bigotimes_{k<j}Z_k,
\end{align}
where $j$ labels the qubit corresponding to the $j$-th orbital ~\cite{jordan1993paulische}. Using this technique, the  Hamiltonian can ultimately be expressed as a sum of Pauli strings, which are the same building blocks we use to construct circuits. Additionally, molecular symmetries can often be exploited to reduce the number of qubits required.

The $\bigotimes_{k<j}Z_k$ term in the JW mapping, which is used to enforce anticommutation by encoding the fermionic parity of all lower-indexed modes, often yields Pauli strings whose Pauli weight scales linearly with system size~\cite{seeley2012bravyi}. This leads to highly non-local terms in the Hamiltonian, which typically compile into long chains of two-qubit entangling gates when implemented on hardware. Alternative mappings such as the Bravyi-Kitaev (BK) and Parity mappings can reduce the Pauli weight of operators from linear to logarithmic in system size~\cite{bravyi_fermionic_2002}. However, it is not always obvious which mapping yields the best practical performance: JW’s local structure can sometimes lead to more hardware-efficient circuits depending on connectivity, error rates, and system size, so the optimal mapping is problem- and architecture-dependent. 

\subsubsection{Variational Quantum Algorithms}
With a qubit Hamiltonian defined, the central task becomes solving for its eigenvalues. The most prominent near-term method for doing so is the Variational Quantum Eigensolver (VQE)~\cite{peruzzo_variational_2014,tilly2022variational}, a hybrid quantum-classical routine in which a quantum computer prepares a trial state using a parametrized ansatz and a classical optimizer iteratively updates the parameters to minimize the measured energy. The success of the VQE method hinges on the choice of an efficient ansatz. While chemically-inspired choices like the Unitary Coupled Cluster Singles Doubles (UCCSD) ansatz are a natural choice~\cite{mullinax2025large}, they often lead to deep circuits. This motivated the development of methods like ADAPT-VQE that systematically grow a compact, problem-specific ansatz iteratively~\cite{grimsley_adaptive_2019}. In contrast with a fixed ansatz, ADAPT-VQE starts with a simple reference state (like the Hartree-Fock state) and incrementally adds operators from a predefined pool. At each step, the algorithm selects the operator that has the largest gradient with respect to the energy, ensuring the most significant contribution to lowering the energy is added next. This process creates a tailored ansatz with fewer parameters and a shallower circuit depth compared to a fixed UCCSD ansatz, making it more resilient to noise on near-term devices.

Crucially, an accurate ground state calculation serves as the foundation for determining molecular excited states. Methods like the Variational Quantum Deflation (VQD)~\cite{higgott_variational_2019} algorithm extend the VQE framework to find higher energy states sequentially: after finding the ground state, VQD finds the next state by running a new VQE with a modified cost function, which penalizes any overlap with the previously-computed ground state. 
The fidelity of these excited state calculations is directly dependent on the quality of the initial ground state simulation, inheriting all of its challenges related to ansatz fidelity, circuit depth, and hardware noise.

\subsubsection{Quantum Phase Estimation and Quantum Imaginary Time Evolution}
While more resource-intensive algorithms like Quantum Phase Estimation (QPE)~\cite{kitaev_quantum_1995} promise a direct path to high-precision energies in the fault-tolerant era by requiring long, coherent quantum evolutions, other approaches involving imaginary time evolution ~\cite{mcardle2019variational,motta_determining_2020} and Krylov subspace methods~\cite{cortes_quantum_2022,yoshioka2025krylov} present compelling alternatives for current hardware. Given a unitary operator $\hat{U}$ and a qubit register containing one of its eigenstates $\ket{\lambda}$, QPE can be used to efficiently approximate $\lambda$, the eigenvalue corresponding to $\ket{\lambda}$, with $p$ bits of precision. QPE also assumes efficient access to controlled-$\hat{U}^{2^j}$ gates, for $0 <j < p -1$.

Imaginary time evolution methods evolve the Schr\"{o}dinger equation in imaginary time, which projects an initial state to the ground state. However, the main challenge with applying imaginary-time evolution on a quantum computer is that the corresponding propagator is not unitary~\cite{nishi2021implementation}. To address this challenge, two main variants have been proposed. Variational Imaginary-Time Evolution (VITE) keeps the state in a parametrized ansatz whose parameters are updated via the McLachlan variational principle~\cite{mcardle2019variational}. Quantum Imaginary-Time Evolution (QITE) instead approximates each short imaginary-time step with a product of local unitary operators chosen by solving a small linear system of measured expectation values, thereby sidestepping the need for a classical optimizer~\cite{motta2020determining}. Outside of ground-state preparation, these algorithms have also found use in partition function estimation, Gibbs-state sampling, and linear partial differential equations~\cite{silva2023fragmented, kumar2024generalising, wang2023symmetry}.

\subsubsection{Quantum Krylov Methods}
Krylov methods work by diagonalizing the Hamiltonian within a small and cleverly-constructed subspace. This subspace is built from a basis generated by repeatedly applying the Hamiltonian to an initial reference state $\ket{\Psi_0}$, creating a basis set like $\left\{\left|\psi_{0}\right\rangle, H\left|\psi_{0}\right\rangle, H^{2}\left|\psi_{0}\right\rangle, \ldots\right\}$. On a quantum computer, the direct application of the Hamiltonian is replaced by time evolution under that Hamiltonian. The algorithm generates a set of basis states by evolving the initial state for different time intervals, $\left|\psi_{k}\right\rangle=e^{-i H t_{k}}\left|\psi_{0}\right\rangle$~\cite{cortes_quantum_2022}. The core of the method then proceeds with a hybrid quantum-classical loop. The quantum computer is used to estimate the matrix elements of the overlap matrix $S_{jk} = \langle \psi_j \vert \psi_k \rangle$ and the Hamiltonian matrix $H_{jk} = \langle \psi_j \vert H \vert \psi_k \rangle$. A classical computer then solves the resulting compact generalized eigenvalue problem, $H v = E S v$, to obtain highly accurate estimates of the ground state and several excited states simultaneously. Since the basis states are discarded immediately after their overlaps are measured, quantum Krylov methods bypass the memory plague of the classical Lanczos algorithm. These methods are generally less resource-intensive than QPE and can be more robust to certain types of errors than VQE, as they do not rely on a complex classical optimization landscape, making them an attractive choice for NISQ hardware.

\subsubsection{Recent Ground State Demonstrations}
Recent ground state benchmark studies now span proof-of-principle calculations to chemically interesting reactions and systematically compare algorithmic and error mitigation strategies across architectures. Using only 2–4 qubits, IBM’s superconducting processors paired a transcorrelated Hamiltonian with an explicitly correlated VQE ansatz to map the H$_2$ and LiH dissociation curves to within approximately 1 mHa accuracy~\cite{dobrautz_ab_2024}. In another notable experiment, Yoshioka \textit{et al.} successfully executed quantum Krylov diagonalization of a 56-site Heisenberg model on a superconducting quantum processor using 43 qubits with shallow circuits~\cite{yoshioka_krylov_2025}. They demonstrated exponential convergence to the ground state energy, a key advantage of quantum Krylov methods. Quantinuum’s H2-2 trapped-ion processor integrated mid-circuit colour-code quantum error correction into QPE and obtained the H$_2$ ground-state within 0.018 Ha, marking the first end-to-end error-corrected chemistry calculation on real hardware~\cite{yamamoto_quantum_2025}. IonQ’s Aria system executed an orbital-optimized pair-correlated (oo-upCCD) VQE with 72 parameters on 12 qubits, mapping full dissociation curves for H$_2$O and BeH$_2$ and retained simulator-level accuracy without explicit error mitigation~\cite{zhao_orbital-optimized_2023}. Beyond qubits, Kim \textit{et al.} used qudit VQE to reach chemical accuracy for H$_2$ and LiH without any error-mitigation overhead~\cite{kim_qudit-based_2024}. Moreover, the open-source BenchQC suite benchmarked VQE on Al$_n^-$ clusters under realistic IBM noise models, finding $<$0.02\% deviation from high-level classical data and providing a reproducible workflow for large-scale, noise-aware algorithm testing~\cite{pollard_benchqc_2025}. 

Since NISQ hardware is inherently noisy, the raw data from a quantum computation are often inaccurate. For today’s NISQ devices, error mitigation, which focuses on post-processing to correct for errors based on noise models, rather than error correction, which focuses on leveraging physical qubits in hardware to correct logical qubits, is crucial. Key error mitigation methods include Zero-Noise Extrapolation (ZNE)~\cite{kandala_error_2019}, in which computations are run at multiple noise levels to extrapolate to a zero-noise result; Probabilistic Error Cancellation (PEC)~\cite{temme_error_2017}, which models and inverts the noise; and Symmetry Verification~\cite{bonet-monroig_low-cost_2018}, which discards results that do not adhere to known chemical symmetries including particle number and spin symmetries. The combination of more efficient simulation algorithms, Hamiltonian reduction techniques, and error-mitigation techniques is steadily pushing the boundaries of what is possible using quantum computers. 

\subsection{Hamiltonian Simulation}
\label{ssec:ham-sim}
Hamiltonian simulation is a central problem in quantum computing and chemistry and serves as a subroutine in many modern quantum algorithms. Assuming a bijective mapping between a physical system and a quantum computer, the simulation problem consists of approximating the result of time evolving a given initial physical state for time $t$. After mapping the initial state to the quantum computer, we then simulate the physical system's dynamics by applying the time-evolution operator. These two steps correspond, respectively, to the state preparation and computation stages of a quantum circuit. The result will be a qubit register---a set of qubits with the same function---that stores an approximation to the time-evolved state. This approximation can then be measured to estimate physical properties or used as input to subsequent quantum computations.

Hamiltonian simulation is important for quantum scientists not only because it is directly related to quantum chemistry and physics, but also because it has been shown to be \textbf{BQP}-complete \cite{manenti2023quantum}. From a complexity theory angle, this means that Hamiltonian simulation, in the general case, is one of the hardest problems we can hope to solve efficiently using a quantum computer. It also means that every problem that can be solved efficiently on a quantum computer can be expressed as a Hamiltonian simulation problem. 

\subsubsection{Trotterization}
As in classical methods, product formulas are a common technique for approximating the time-evolution operator. Given a Hamiltonian $\hat{H}$ expressed as the sum of $P = poly(n)$ $k$-local terms and its associated time-evolution operator $\hat{U}_t$, a first-order Trotter formula can be used to generate an approximation $\hat{V}_t$:

\begin{equation}
    \hat{H} = \sum_{j=0}^P \hat{H}_j \quad \longrightarrow \quad \hat{U}_t = e^{-i\left(\sum_{j=0}^P \hat{H}_j\right)t} \approx \prod_{j=0}^P e^{-i\hat{H}_j t} = \hat{V}_t
\end{equation}
with respect to time $t$ and accuracy $\epsilon$, defined as $\epsilon > \|\hat{V}_t -
\hat{U}_t\|_o$ where $\|\cdot\|_o$ denotes the operator norm \cite{Lloyd1073}. First-order Trotterization requires an
$O(t^2/\epsilon)$ circuit depth. Higher-order approximations are also possible and can 
yield more favorable asymptotic complexity, such as the symmetrized second-order decomposition, 
which scales as $O(t^{3/2}/\epsilon^{1/2})$ \cite{suzuki1991general, childs2019faster, berry2007efficient, wiebe2010higher}.
However, it is important to note that asymptotic complexity is not always a reliable proxy for implementability and higher-order Trotter formulas typically require deeper circuits. 

\subsubsection{Linear Combination of Unitaries}
More recent methods approximate the time evolution operator by making use of the Linear Combination of Unitaries (LCU) algorithm \cite{childs2012hamiltonian}. LCU enables the encoding of the sum of unitary operators, which normally is non-unitary, into a unitary matrix acting on a larger system. If our matrix of interest can be expressed as
\begin{equation}
    \hat{H} = \sum_{j=0}^P\alpha_j\hat{U}_j, 
\end{equation}
where all $\hat{U}_j$ are unitaries acting on $n$ qubits, then we can construct a block encoding $\hat{B}$ with $\alpha = \sum_{j=0}^J \alpha_j > 0$ such that
\begin{equation}\label{eq:block-encoding}
    \bra{k}\hat{H}\ket{l} = \alpha\bra{0}^p\!\bra{k}\hat{B}\ket{0}^p\!\ket{l},
\end{equation}
where $\ket{k}$ and $\ket{l}$ are two $n$-qubit generic basis elements and $\ket{0}^p$ represents a register of at most $\lceil \log_2 P \rceil = p$ ancilla qubits set to the $\ket{0}$ state. The result is a unitary operator that acts on $n\!+\!p$ qubits with the net effect of applying $\hat{H}$ on the $n$ qubit register. The cost of applying LCU scales as $O(nPp)$. The successful application of $\hat{B}$ can also be conditioned on measuring $\ket{0}^p$ on the ancilla register for better accuracy, although this can lead to a higher complexity measurement scheme. 

\subsubsection{Quantum Signal Processing and Qubitization}
Modern algorithms based on Quantum Signal Processing (QSP) and qubitization \cite{Low_2017,Gily_n_2019,motlagh2024generalized} provide another way to perform general-purpose quantum computation from the lens of functional approximation. 
The idea is that, given a Hermitian matrix $A$ ($||A|| < 1$) encoded inside a unitary
\[
U = \begin{pmatrix} A & \phantom{-}*\phantom{\,} \\ * & \phantom{-}*\phantom{\,} \end{pmatrix}
\]
such that $A = \Pi U \Pi $ with $\Pi$ a projector that flags the location of $A$, QSP provides a constructive way to realize a degree-$d$ polynomial transformation on $A$ as $P_d(A)$ by querying $U$ only $d$ times,
\begin{align}
    e^{i \phi_0 \Pi} \left[ \prod_{k=1}^d U e^{i \phi_k \Pi} \right]  = \begin{pmatrix} P_d(A) & \phantom{-}* \phantom{\,} \\ * & \phantom{-}*\phantom{\,} \end{pmatrix}. \label{eq:qsp}
\end{align}
Here, $U$ is called a \emph{block-encoding} of $A$, and $\{\phi_0, \phi_1, \cdots, \phi_d \}$ is a collection of phase angles that determines the shape of $P_d(\cdot)$. The polynomial $P_d(\cdot)$ can be chosen to approximate any analytical functions $f(\cdot)$ with only very mild constraints. This provides a unified way to realize many known quantum algorithms \cite{martyn2021grand}. Note that we assume $A$ to be Hermitian here for simplicity of notation, and this constraint can be lifted \cite{Gily_n_2019}.

For Hamiltonian simulation, let $A = \hat{H} / \alpha $ and $f(x) = e^{-i x t}$, where $\alpha$ is a scaling factor that guarantees that the norm of $\hat{H}/\alpha$ is less than 1. Then, QSP can find an optimal degree-$d$ polynomial $P_d(\hat{H} / \alpha)$ that approximates $e^{-i \hat{H} t}$ \cite{Low_2017,berry2024doubling}, where $d \sim O\left(\alpha t  + \frac{\log(\nicefrac{1}{\epsilon})}{\log \log (\nicefrac{1}{\epsilon})}\right)$ and $\epsilon$ is the simulation error, namely the distance between $P_d(x)$ and $e^{-ix t}$. When $\hat{H}$ is an electronic or electron-nuclear Hamiltonian mapped to qubits, the QSP circuit in Eq. \eqref{eq:qsp} provides a way to simulate electronic or chemical reaction dynamics.

\section{Developments in Quantum Computation for Quantum Chemistry Beyond the Ground State} \label{quantumchemnoground}

\subsection{Reaction Mechanisms and Molecular Dynamics in the Born-Oppenheimer Approximation}

One of the central questions underpinning all of chemistry is: \textit{how does a chemical reaction proceed?} Within the Born-Oppenheimer approximation---which separates nuclear and electronic motions---reaction mechanisms are typically determined using classical computational methods. These fall into two main categories. The first involves explicit reaction pathway construction, using methods such as the string method~\cite{weinan2002string}, the dimer method~\cite{henkelman1999dimer}, Transition Path Sampling (TPS)~\cite{dellago2002transition}, or the Nudged Elastic Band (NEB) method~\cite{henkelman2000climbing}. The second category encompasses reactive molecular dynamics techniques that sample reaction trajectories to infer mechanisms~\cite{van2001reaxff}. Both approaches require computing all or parts of the potential energy surface (PES), which describes how energy changes with atomic positions, using electronic energies, along with forces (first-order gradients) and Hessians (second-order gradients). Forces and Hessians help identify key points along pathways, including minima (reactants, products, or intermediates) and saddle points (transition states). In dynamics calculations, forces determine particle accelerations to advance trajectories.

Although classical electronic structure methods, like Density Functional Theory (DFT) and Coupled Cluster (CC), may suffice for determining the energies of stable structures that can be described by a single electron configuration (few Slater determinants), they often struggle to describe transition states and bond breaking/forming regions. These regions involve strong electron correlation, requiring more accurate but computationally more expensive multireference methods~\cite{holmes2016heat,motta2018ab,foulkes2001quantum}. This challenge makes quantum computation an attractive alternative, as it could handle these correlations more efficiently and at lower cost. Nonetheless, quantum-based reaction pathway or dynamics calculations generally need not just energies, but also energy gradients—which are challenging, but feasible to compute on quantum hardware~\cite{o2022efficient}.

Recent developments in quantum algorithms and NISQ-era hardware are beginning to translate these theoretical advantages into practical applications for chemical simulations~\cite{zhang_quantum_2025}.
Although hardware demonstrations are still confined to modest system sizes, they are growing rapidly and provide early proof-of-concepts for reaction pathway mapping. For example, Google's 2020 Sycamore experiment modeled two competing isomerization pathways of diazene N$_2$H$_2$~\cite{google_ai_quantum_and_collaborators_hartree-fock_2020}. After freezing two core orbitals, the problem was reduced to 10 active qubits embedded in a 12-qubit Sycamore line; each geometry was prepared with a basis-rotation circuit containing 50 $\sqrt{i\mathrm{SWAP}}$ and 80 $R_z$ gates. This was essentially a Hartree-Fock calculation at several molecular geometries, run as a hybrid quantum-classical loop using the Variational Quantum Eigensolver (VQE) algorithm. Utilizing error mitigation techniques like occupation number post-selection and McWeeny purification improved the calculation's fidelity to $>0.98$ and predicted correct transition state ordering with an energy gap of $41 \pm 6 \ mE_h$ versus the true gap of $40.2 \ mE_h$. This study therefore doubled the qubit count of IBM’s earlier six-qubit BeH$_2$ profile~\cite{kandala_hardware-efficient_2017} while showing that aggressive error-mitigation plus parameter transfer can deliver chemically-meaningful reaction energetics on present-day hardware. It also served as one of the first demonstrations of mapping a full reaction pathway, in this case geometry-by-geometry, on quantum hardware. 

While effective, mapping pathways geometry-by-geometry as in the Sycamore experiment has limitations, such as restarting VQE from scratch for each geometry, which can lead to potentially unnecessary overhead. One recent approach that attempts to tackle the restarting issue is a smooth-geometry variational algorithm, GeoQAE, which follows the ground state adiabatically as bonds break and form~\cite{yu_geometric_2022}. This approach involves preparing a ground state wavefunction at an easy, near-equilibrium geometry and subsequently evolving the system along a discretized nuclear path. At each step, the Hamiltonian is smoothly interpolated so that only small geometric changes are made, maintaining the system in the ground state. This approach avoids the fresh quantum computation of the electronic energy for each geometry by reusing the wavefunction from the prior calculation. GeoQAE reproduced the potential energy surface of the $H_2 + D_2 \rightarrow 2 HD$ reaction on an 8-qubit Hamiltonian with an energy difference of $\sim 10^{-4}$ Ha relative to exact results across configurations. 

A key challenge uniting these examples is computational cost. Even if the same circuit ansatz suffices at every geometry, the circuit must be executed $N_{\text{geom}}$ times, often requiring deeper circuits near transition states and bond-breaking regions. As a rule of thumb, mapping a full reaction path demands one to two orders of magnitude more shots and gate operations than a single-point ground-state calculation. However, algorithms that reuse the wavefunction smoothly across geometries, combined with error mitigation, are beginning to reduce that overhead to within experimental reach, setting the stage for extending these methods to dynamical simulations. 

With static reaction profiles now accessible on today's hardware, the next frontier is to predict the time-dependent evolution of nuclear (or coupled electron-nuclear) wavefunctions, shifting the focus from energetics to dynamics. One key approach in this direction is Born-Oppenheimer molecular dynamics (BOMD), in which quantum mechanics is used to compute the potential energy and forces for a molecular configuration, which are then used to iteratively update the next configuration~\cite{mouvet2022recent}. Like the reaction pathway techniques above, BOMD depends on an accurate potential energy surface, but challenges in computing energy gradients directly on quantum hardware have led to hybrid quantum-classical methods, where quantum computers handle energies, while classical computers compute gradients and geometry updates. Early efforts used finite-difference or surrogate methods for forces on classical computers~\cite{iyer2024force,gustafson2024surrogate,o2019calculating}. More recently, quantum-computed energies have been used to train machine-learned potentials (MLPs) for efficient dynamics simulations~\cite{khan2024quantum,huang2022machine}, such as via transfer learning to refine models trained on classical Density Functional Theory (DFT) data~\cite{khan2024quantum}. These have enabled quantum-informed force fields for systems like water and small biomolecules. However, these approaches assume that quantum-computed energies can be obtained more accurately and efficiently than on classical hardware to make their computation worthwhile, a goal not yet achieved. 

\subsection{Non-Born Oppenheimer Molecular Dynamics}
Building on the previous section's discussion of reaction mechanisms and BOMD, where nuclear motion is treated classically on quantum-computed potential energy surfaces, we now turn to more advanced approaches that incorporate quantum effects into nuclear dynamics. This includes quantum molecular dynamics, which in general, can occur in the Born-Oppenheimer (BO) approximation or beyond it, in non-BO regimes. Non-BO dynamics, in particular, evolves the full wavefunction describing both electrons and nuclei according to the time-dependent Schr\"{o}dinger Equation, offering greater accuracy for processes like coupled proton-electron transfer, in which nuclear quantum effects are significant. 
This accuracy is accompanied by a significantly greater computational expense: mixed quantum-classical methods that require the computation of multiple electronic states and their couplings are roughly one order of magnitude more expensive than BOMD~\cite{subotnik2016understanding}, while fully quantum methods such as Multi-Configurational Time-Dependent Hartree~\cite{beck2000multiconfiguration} and quantum wave packet techniques can be multiple orders of magnitude more demanding. This expense highlights a key opportunity for quantum computing to achieve potential speedups, especially in fully quantum simulations.

Early work demonstrated that coupled electron-nuclear non-BO dynamics can be simulated in polynomial time on a digital quantum computer using split-operator, real-time propagation~\cite{kassal2008polynomial}. This work propagated a coupled electron-nuclear wavefunction and demonstrated that reaction observables like state-to-state transition probabilities and thermal reaction rates can be obtained via quantum measurements. The coupled wavefunction was stored on real-space grids using $n$ qubits per Cartesian coordinate plus $4m$ ancilla registers, where $m$ is the number of bits of numerical accuracy, so a $B$-particle system needs $n(3B-6)+4m$ qubits. The evaluation of the pair-wise potential at each time slice costs $\left( \frac{75}{4}m^3 + \frac{51}{2}m^2 \right)$ elementary gates per particle pair, keeping the scaling efficient. Recent algorithms have built on these foundations, often using time-dependent Hamiltonian simulation approaches. Ollitrault \textit{et al.}, for example, proposed a first quantized algorithm for fast nonadiabatic dynamics in which a nuclear wavepacket on two coupled potential energy surfaces (a Marcus model) evolves in time~\cite{ollitrault_nonadiabatic_2020}. They demonstrated that this scheme has a depth that scales polynomially with system size due to the efficient encoding of position coordinates and electronic populations. In 2024, Kale and Kais introduced a quantum algorithm to calculate scattering matrix (S-matrix) elements for chemical reactions~\cite{kale_simulation_2024} through a Hadamard test evaluation of time correlation functions: an $N=256$ grid required just 8 qubits, which store and manipulate the quantum state and an ancillary qubit that stores information about the correlation function; the sample complexity scales as O(1/$\epsilon^2$). They demonstrated the one-dimensional collinear hydrogen exchange reaction $H+H_2 \rightarrow H_2 + H$ as a proof of concept, which should fit comfortably on $\sim9$ qubits with a depth dominated by standard Trotterized propagators. These methods illustrate how quantum algorithms can capture quantum nuclear effects---like tunneling or zero-point energy---that classical dynamics often miss, potentially enabling more accurate predictions of reaction rates in complex environments.

Bosonic quantum devices, which represent an alternative hardware paradigm, provide an efficient way to simulate non-BO dynamics by natively representing vibrational modes without the need for qubit mappings~\cite{dutta2024chemistry,vu2025computational}. Izmaylov and coworkers~\cite{malpathak_simulating_2025}, for instance, introduced a framework for digital quantum simulation of vibrational dynamics on bosonic devices. Their approach partitions the vibrational Hamiltonian into solvable anharmonic fragment Hamiltonians that can be propagated with native Kerr or cross-Kerr gates on current bosonic hardware. A single trotter step for an $N$-mode system then requires only $N_f N(N+1)/2$ non-Gaussian gates, where $N_f$ is the number of bosonic fragments, which is four to five orders of magnitude fewer gates than the T-gates demanded by a fully-commutating Pauli decomposition on qubits. For small molecules (CO, H$_2$O, H$_2$S, CO$_2$), the scheme identifies and requires just 2-7 solvable fragments (versus 54-170 Pauli groups) and reproduces the lowest four vibrational levels to $< 1\ cm^{-1}$. To validate dynamics, they track coherent proton tunneling in a two-dimensional double-well using just four fragments while keeping overlap errors below $10^{-3}$ for Trotter steps $\Delta t < 10^{-2} au$. This bosonic approach reduces resource overhead, demonstrating how quantum hardware can directly mimic molecular vibrations, a key quantum nuclear effect in reactions, at reduced cost.

Beyond digital gate-based proofs-of-concept, a series of analog and hybrid experiments have started to capture non-BO chemical dynamics on hardware. In one key example~\cite{navickas_experimental_2025}, a trapped-ion mixed-qubit-bosonic simulator encoded the electronic states of the single $^{171}$Yb$^{+}$ ion and its harmonic nuclear vibrations. This setup was able to reproduce ultrafast, non-adiabatic wavepacket splitting at conical intersections for photoexcited allene, butratriene, and pyrazine. Femtosecond population transfer was achieved with just one ion; a purely digital simulation would need 11 qubits and $\sim 10^5$ CNOT gates. Photonic continuous-variable processors have likewise been used to mimic vibronic energy transport in molecules, leveraging bosonic hardware to bypass large qubit overheads~\cite{zhu_large-scale_2024}. Complementing these efforts, Google's analogue-digital quantum simulator~\cite{andersen_thermalization_2025} has pushed hybrid simulation to the 70-qubit scale. While not a direct simulation of a chemical reaction, this work is highly relevant as it showcases the power of a hybrid architecture. By combining the programmability of digital gates with the efficiency of analogue evolution on a superconducting-qubit processor, the researchers leveraged a 69-qubit Sycamore-class processor whose tunable-coupler lattice natively realizes a $U(1)$-symmetric 2D XY Hamiltonian with cycle errors below $10^{-3}$. By interleaving high-fidelity analog evolution with universal one- and two-qubit gates, they reached the Porter-Thomas scrambling regime within $<60$ ns, tracked coarsening across the Kosterlitz-Thouless transition, and measured Renyi entropies for subsystems up to 12 qubits. The same hybrid recipe---global, boson-like interaction plus digital rotations---maps naturally onto coupled electron-nuclear or vibronic models, indicating that superconducting analog-digital platforms may be poised to tackle challenging non-BO dynamics at the $\sim10$ qubit scale. These experiments highlight that analog elements can accelerate simulations by exploiting natural hardware dynamics, reducing the gate counts that limit digital approaches.

The field of quantum simulation for reaction dynamics is advancing rapidly by exploring a range of systems, from benchmark reactions like hydrogen exchange to vibrational spectra of molecules like CO and H$_2$O. Progress is being driven by algorithmic advancements such as time-dependent Hamiltonian simulation and Hamiltonian fragmentation. A notable trend is the strategic choice between traditional qubit-based methods and emerging bosonic quantum devices, which offer an efficient alternative by avoiding the overhead of the boson-qubit mapping. This is impactful for resource optimization where the focus is on minimizing the number of gates and complexity. Diverse platforms—superconducting qubits, trapped ions, and photonic systems—are addressing challenges by capitalizing on each technology's strengths. As these tools mature, they hold promise for transformative insights into non-BO processes, from photochemistry to quantum tunneling in biological systems.

\subsection{Electron Dynamics}
For reasons similar to those for non-BO dynamics, significant quantum speedups have been proven to be possible for the simulation of electronic dynamics under both 1st and 2nd quantization on quantum computers. In electron dynamics, one studies the evolution of electrons and their corresponding wave functions through a chemical system often after an initial laser or other excitation. While, at a fundamental level, modeling electron dynamics necessitates solving the time-dependent Schr\"{o}dinger Equation, modeling dynamics with high-accuracy, especially for time-dependent Hamiltonians and in the presence of electron correlation, has been historically difficult. Many approximations have been developed, such as mean-field approximations like the Redfield Equation, but getting precise time dynamics can prove exceedingly costly, requiring the solution of hierarchical sets of coupled equations or exact propagation. The need to model full quantum dynamics leaves room for quantum computation to provide speedups. 

\subsubsection{Non-Interacting Free Fermions}
For non-interacting free fermions, it was shown in Ref. \cite{stroeks2024solving} that a $polylog(n)$ size circuit can be constructed to block-encode $n$-orbital free-fermionic Hamiltonians with sparse one-electron integrals. This circuit can be combined with QSP to give rise to exponential speedups on quantum hardware relative to classical hardware for electron dynamics simulations of certain free fermion systems,  providing the 1-RDM of the initial state is also sparse. Circuit compression techniques based on Lie algebra have also been developed that can achieve linear-depth Trotter simulation (controlled or uncontrolled) of free-fermions in second quantization with long-range hopping (or, equivalently, on arbitrary lattices) \cite{kokcu2023algebraic}. Experimental demonstrations have been performed for a $4\times 4$ tight-binding model on the \textit{ibmq\_washington} and Quantinuum H1-1 trapped-ion quantum computers.

\subsubsection{Interacting Electrons in First Quantization}
The situation is much more complicated for interacting electrons. Early works \cite{lidar1999calculating,kassal2008polynomial} showed that an exponential speedup is possible in first quantization by simply performing Trotter time-evolution of the kinetic and potential operators interpreted by a quantum Fourier transform, which effectively makes the kinetic and potential operators both diagonal, dramatically speeding up the simulation. A total of $n(3B-6)+4m$ qubits are needed to simulate a $B$-particle system on $2^n$ grid points using $2^m$ points to discretize the Coulomb potential. The gate count for each Trotter time step is roughly $B^2 (\frac{75}{4} m^3 + \frac{51}{2} m^2)$. While the original method in Ref.~\cite{kassal2008polynomial} was proposed for coupled electron-nuclear simulations, it applies to electronic dynamics as well. With more recent developments in Trotter error analysis and scaling \cite{low2023complexity,childs2021theory,su2021nearly}, Ref.~\cite{babbush2023quantum} presented a tightened Trotter gate count of $O(n^{1/3} \eta^{7/3} t + n^{2/3} \eta^{4/3} t) \left( \frac{nt}{\epsilon}\right)^{o(1)}$, demonstrating that exact interacting electronic dynamics simulations on a quantum computer can exhibit a quartic speedup over the cost of mean-field dynamics on classical computers $O(n^{4/3} \eta^{7/3} t + n^{5/3} \eta^{4/3} t) \left( \frac{nt}{\epsilon}\right)^{o(1)}$. We also note a broad class of randomized product formulas such as qDRIFT \cite{Campbell_2019} and its high-order generalization -- qSWIFT \cite{nakaji2024qswift}. These randomized simulation algorithms can often improve the gate count as compared to the deterministic product formulas by using different Trotter decompositions for each Trotter step, such that the overall Trotter error can be canceled to effectively higher order. New Hamiltonian approximation techniques including stochastic sparsification \cite{ouyang2020compilation} have been combined with these randomized simulation algorithms to improve their performance. 

\subsubsection{Initial State Preparation in First Quantization} One of the important issues in the first-quantized simulation of electronic dynamics is the preparation of an initial state that satisfies fermionic statistics, i.e., the total wave function has to change sign under a permutation of two electrons. Refs. \cite{PhysRevA.106.032428,su2021fault} constructed such anti-symmetrized Slater determinants with a gate cost of $O(\eta^2 n \log(n))$ based on Givens rotation \cite{PhysRevLett.120.110501}. Ref. \cite{babbush2023quantum} improves the gate cost to $\tilde{O}(n \eta)$ based on prior anti-symmetrization works \cite{berry2018improved}. The efficient anti-symmetrization technique proposed by these works came at the cost of ignoring the spin part of the electronic wave function and was also restricted to a single Slater determinant, which is of limited use for strongly correlated systems with multireference character and the spin-orbit coupling often observed in transition metal complexes \cite{schaffer2016recent} and lanthanide and actinide chemistry \cite{cotton2024lanthanide}. Based on a novel group theoretical approach that unifies the treatment of all finite symmetries in quantum simulation, Ref. \cite{bastidas2025unification} overcame these challenges by providing a way to prepare anti-symmetrized, \emph{correlated, and spinful} electronic wave functions in first-quantization, as demonstrated on real quantum hardware for the H$_2$ molecule in the STO-3G basis. This method also allowed parallel quantum simulation of multiple symmetric sectors in one-go. Such techniques may be transferable and offer promise across the problem classes discussed in this review.

\subsubsection{The Interaction Picture, Time-Dependent Hamiltonian Simulation, and Beyond} 
Ref. \cite{low2018hamiltonian,berry2019timedepedentHS} proposed new time-dependent Hamiltonian simulation algorithms based on a truncated Dyson series. Ref. \cite{low2018hamiltonian} moreover showed that it is possible to first transform the Hamiltonian into the interaction picture and then use the truncated Dyson series to perform the time-dependent Hamiltonian simulation in the interaction picture. Several more recent works also proposed various Hamiltonian simulation techniques in the interaction picture \cite{bosse2025efficient,sharma2024hamiltonian,fang2025time,An2022timedependent}, but the applications to quantum chemistry have yet to be developed.

Based on Ref.~\cite{low2018hamiltonian}, Ref.~\cite{babbush2019quantum} showed that a sublinear circuit depth $O(n^{1/3} \eta^{8/3})$ for basis set size $n$ is possible by simulating electronic dynamics in the plane wave basis in the interaction picture. 
Upon moving into the rotated frame of the kinetic operator, the original potential term becomes a time-dependent Hamiltonian that has a smaller norm than the kinetic term. Recall (see Sec. \ref{ssec:ham-sim}) that the simulation cost depends on the norm of the Hamiltonian, so this interaction picture transformation effectively reduces the norm of the total Hamiltonian. 
The resulting interaction picture Hamiltonian is then Trotterized into small time steps; within each time step, the Hamiltonian is taken as approximately  constant. For each Trotter step simulation, a Dyson series expansion of the unitary dynamics is performed and truncated to a certain order \cite{2015Berry}. This sum over different series is then performed using the linear combination of unitary (LCU) algorithm. The algorithm relies on the block-encoding of the interaction picture Hamiltonians as well as amplitude amplification. The optimized algorithm for block-encoding and explicit circuits was analyzed in Ref.~\cite{su2021fault}, which also showed that scaling similar to that presented in Ref.~\cite{babbush2019quantum} is possible using a real-space grid basis. Ref.~\cite{su2021fault} also analyzed explicit gate counts needed to perform quantum phase estimation on realistic chemical systems including ethylene carbonate and LiPF$_6$, and compared with prior art. 

\subsubsection{Interacting Electrons in Second Quantization}
In second-quantization, Ref.~\cite{low2023complexity} proposed a low-rank recursive block encoding strategy to implement a single Trotter step using qubitization, and then multiplied all Trotterized steps together. This gives an improved gate count for simulating the uniform electron gas as $O((n^{5/3} / \eta^{2/3}  + n^{4/3} \eta^{1/3}  ) n^{o(1)})$.

In these qubitization-based algorithms, the final unitary evolution of interest can only be achieved with a finite success probability due to the use of ancillary qubits in the block-encoding. As a result, various versions of amplitude amplification (AA) algorithms have to be used to boost the success probability to unity, which introduces additional overhead. Ref.~\cite{martyn2023efficient} circumvented this issue by developing an efficient and fully coherent algorithm without amplitude amplification to perform Hamiltonian simulation that achieves an additive query scaling of $O\left( || H || t + \log(\frac{1}{\epsilon} + \log\left(\frac{1}{\delta} \right) \right)$, where $\delta$ is the failure probability. The “one-shot” algorithm in Ref. \cite{martyn2023efficient} is significantly better than naive AA and also over-performs QSP + robust, oblivious AA for long simulation time $t$ and low to intermediate simulation error $\epsilon$. The algorithm has been validated on small systems, including the Heisenberg dimer under both time-independent and time-dependent external fields, as well as the femto-second charge oscillator dynamics of the H$_2$ (STO-3G basis, second-quantization) molecule by performing numerical simulation of the explicit quantum circuits. Similar to Trotter, randomized versions of QSP-based algorithms have been proposed that mix different polynomials, which halves the overall cost of Hamiltonian simulation \cite{martyn2025halving,wang2024faster}. 

We note that simulation methods that combine different simulation techniques together \cite{hagan2023composite} have demonstrated improved performance for simple systems including the jellium model, H$_3$ molecule, and Heisenberg spin models \cite{PhysRevResearch.6.013224}.

\subsubsection{Summary}
As one of the most promising quantum chemistry applications of quantum computers beyond the ground state, we emphasize the need for developing quantum computing methods that can simulate spinful electronic dynamics, possibly with relativistic effects~\cite{PhysRevA.85.030304}. While our review mostly focused on asymptotic scaling, there are concrete gate counts estimated for many of the algorithms in the above (for example, Ref.~\cite{nakaji2024qswift}). We expect a combination of circuit optimization, faithful algorithmic approximation, and steady progress on fault-tolerant hardware development in the near future to push the limit of these Hamiltonian simulation algorithms to ultimately realize practical quantum hardware execution and utility.

\subsection{Finite Temperature Quantum Chemistry}
While most quantum chemistry focuses on the ground state because, at most temperatures, electrons reside in their ground state, there are some situations in which the temperature is sufficiently high to excite electrons into an ensemble of higher-lying excited states. In these situations, which include molecules and materials at high pressures and temperatures~\cite{mcmahon2012properties,liu2018ab}, one must instead rely on finite temperature quantum chemistry, whose objective is to obtain a system's ensemble of states. As with ground state electronic structure problems, finite temperature problems can again be classified into static and dynamic problems. Quantum simulation needs to capture these non-unitary processes, which is at inherent odds with the unitary dynamics quantum computers are designed to perform. Thus, additional techniques have been developed to tackle this problem.

One of the primary objectives for static problems is to prepare a mixed quantum state of a target system of interest at a given temperature for a given ensemble. Most recent quantum computing works focus on preparing equilibrium states in the canonical ensemble, i.e., the Gibbs state. How to prepare Gibbs states for local and quasi-local Hamiltonians for spins \cite{10756136,rouz2024optimal} and fermions \cite{ramkumar2025high} has been demonstrated from the quantum information perspective. Quantum information analyses have been performed on the high- to infinite-temperature limit, suggesting that, at infinite temperature, fermionic states are simple mixtures of Gaussians, which can be adequately described on classical computers efficiently. These theoretical proofs agree with extensive data from previous classical quantum Monte Carlo studies of finite temperature electronic structure in the high-T limit, in which it was found that the sign or phase problem is only very mild and exact observables can be obtained efficiently~\cite{liu2018ab,motta2018ab,lee2022twenty}. 
Fundamental results for intermediate and low temperatures are more challenging to obtain, but tensor network methods can often provide a practical way to probe the hardness of these states via the bond dimension~\cite{cocchiarella2025lowtemperaturegibbsstatestensor}. These infinite to high temperature results are interesting  in that they reveal fundamental complexity, but are less useful for practical chemical problems.

On the practical side, the preparation of mixed states requires the execution of non-unitary processes on quantum computers. One way is to rewrite the non-unitary operators as a classical ensemble average of many unitary operators where each unitary can be performed on a quantum computer. The statistical average is often performed by sampling the quantum computer via measurements. Examples of this approach include quantum Metropolis sampling~\cite{temme2011quantum}, the more recent Markov-chain Monte Carlo with sampled pairs of unitaries (MCMC-SPU) method~\cite{matsumoto2025quantum}, and other Gibbs samplers~\cite{kastoryano2016quantum,chen2023quantum,Gu_2024}. We refer readers to Table I of Ref. \cite{chen2023quantum} for a comparison of some existing thermal state preparation methods with provable accuracy guarantees. Ref.~\cite{Gu_2024} provides classical emulation results for the LiH (8 qubits) and H$_2$O (12 qubits) molecules at both zero and finite-temperature. These algorithms often achieve a tradeoff between quantum circuit depth and sample complexity.
Interestingly, for short imaginary-time evolution, the non-unitary operator can be well approximated by a unitary operator (up to a normalization constant). This, combined with Trotterization, leads to the quantum imaginary-time evolution (QITE) method \cite{motta2020determining}, where prototype circuits on the Rigetti quantum
virtual machine and Aspen-1 quantum processing unit were executed to demonstrate finite temperature state preparation on 1- and 2-qubit systems. QITE was also used to calculate the finite-temperature static and dynamical properties of larger spin systems with up to four sites on five-qubit IBM quantum devices~\cite{sun2021quantum}.

An alternative way to approach non-unitary dynamics is to cast mixed states into pure states or turn a non-unitary process into unitary ones. Then, the finite temperature problems can be reduced to zero-temperature state preparation and quantum dynamics problems (Sec.~\ref{ssec:ham-sim}), followed by tracing out or performing a projective measurement on part of the qubits. As compared to the ensemble average, these methods effectively perform the average quantumly using additional ancillary qubits. By increasing the dimensionality, the mixed state is \emph{purified} into a pure state and the non-unitary operator is \emph{dilated} into unitary ones. 

\begin{sidewaystable}
\centering
\tiny
\caption{Asymptotic time-complexity estimates for representative quantum algorithms for chemistry beyond ground state. Each entry lists an algorithm, its reported computational scaling, a reference, and the system(s) to which it was applied (if applicable). Variables are defined in the main text.}
\renewcommand{\arraystretch}{.8}
\setlength{\tabcolsep}{8pt}
\begin{tabular}{p{2.5cm}p{3.5cm}l>{\raggedright\arraybackslash}p{3cm}l}
\toprule
\textbf{Category} & \textbf{Algorithm} & \textbf{Scaling} & \textbf{System (qubits)} &  \textbf{Ref.}\\ 
\midrule
Reaction Mechanisms & VQE & $O(n^{4})$ per geometry & $BeH_2$(6), Diazene(12)  & ~\cite{google_ai_quantum_and_collaborators_hartree-fock_2020, kandala_hardware-efficient_2017}\\ % n is no. of orbitals
 \& BO Dynamics & GeoQAE & $O(n^{4})$ with a small prefactor & $H_2 + D_2 \rightarrow 2 HD$ (8) & ~\cite{yu_geometric_2022} \\
\midrule
Non-BO Dynamics & Split operator (first quantization) & \(O(B^2 (\frac{75}{4} m^3 + \frac{51}{2} m^2))\) & B-particle system ($n(3B-6)+4m$)  & ~\cite{kassal2008polynomial} \\
& Lie-Trotter-Suzuki product formulas & $O(poly(N))$ & Two-surface Marcus Model (18)  & ~\cite{ollitrault_nonadiabatic_2020} \\ % N=number of grid points 
& $S$-matrix via Hadamard test & $O(poly(N)) \times O(1/\epsilon^2)$ & $H+H_2$ reaction(9) & ~\cite{kale_simulation_2024} \\ % N=number of grid points 
& Bosonic fragmentation with Kerr and cross-Kerr interactions & $O\!\bigl(N_{f}N(N+1)/2\bigr)$ non-Gaussian gates &CO (4 q / 1 mode); H$_2$O (9 q / 3); H$_2$S (9 q / 3); CO$_2$ (12 q / 4) & ~\cite{malpathak_simulating_2025}\\ % N_f is the no. of bosonic fragments and N is number of modes/DOF
& Analog trapped-ion simulation & Single laser pulse per step (digital equiv.\ $\sim10^{5}$ CNOTs) & photoexcited allene, butratriene, and pyrazine (1 $^{171}$Yb$^{+}$ qudit + 2 phonon modes) [digital equiv. 11 qubits] & ~\cite{navickas_experimental_2025}\\ % 
\midrule
Electron Dynamics  & Qubitization, QSP, QSVT& \(\Theta(||H|| ~|t| + \log(1/\epsilon) + \log(1/\delta) \) & H$_2$ (4), Heisenberg dimer (2) & ~\cite{martyn2023efficient} \\
 & Qubitization, QSP, QSVT& \( polylog(n) \) & some free fermion Hamiltonians & ~\cite{stroeks2024solving} \\
& Trotter \& product formulas & \(O(n^2)\) & arbitrary free fermion Hamiltonians  & ~\cite{kokcu2023algebraic} \\
& Trotter \& product formulas & \(O(B^2 (\frac{75}{4} m^3 + \frac{51}{2} m^2))\) & electron-nuclei in 1st-quantization  & ~\cite{kassal2008polynomial, lidar1999calculating}\\
& Trotter \& product formulas & $O(n^{1/3} \eta^{7/3} t + n^{2/3} \eta^{4/3} t) \left( \frac{nt}{\epsilon}\right)^{o(1)}$ & generic interacting electrons  & ~\cite{su2021nearly, babbush2023quantum} \\
& Trotter \& product formulas & $O((n^{5/3} / \eta^{2/3}  + n^{4/3} \eta^{1/3}  ) n^{o(1)})$ & uniform electron gas, 2nd quantization  & ~\cite{low2023complexity} \\
& Randomized Trotter, qSWIFT & $10^{11} \sim 10^{13}$ gates for $t = 100$ Hartree$^{-1}$, $\epsilon = 0.001$  & propane (STO-3G, 46 qubits), ethane (6-31G, 60 qubits), carbon
dioxide (6-31G, 54 qubits) & ~\cite{nakaji2024qswift} \\
\midrule
Other Ham Sim Algorithms& qDRIFT & $O(\alpha^2 n^2 t^2)$ & $n$ qubits  & ~\cite{Campbell_2019} \\
 & Interaction Picture (Dyson series) & $\tilde{O}(\alpha n^2 t)$ & $n$ qubits & 
~\cite{low2018hamiltonian} \\
& Interaction Picture (Magnus expansion) & $O((nt)^\gamma (\alpha n t)^{1 + o(1)})$ & $n$ qubits & 
~\cite{sharma2024hamiltonian} \\
\midrule
Finite Temperature & Quantum Metropolis Sampliing (QMS) & \(O(\frac{\mathrm{poly}(n)}{\Delta})\); $\Delta$ = spectral gap & n qubits  & ~\cite{temme2011quantum}\\
Chemistry & QAOA-inspired TFD state preparation & \(O(p n)\); $p$ \text{layers}; $n$ qubits & Transverse-Field Ising Model (TFIM) & ~\cite{zhu2020generation} \\
& MCMC‑SPU & \(O(\exp(2\beta))\) number of trials & 1D TFIM  & ~\cite{matsumoto2025quantum} \\
& Clifford+$k$Rz & depends on accuracy and truncation of Clifford hierarchy  & LiH (8 qubits), H$_2$O (12 qubits)  & ~\cite{Gu_2024} \\
& Hybrid VQE + AIMD & depends on VQE ansatz and accuracy & H$_2$, H$_3^+$  & ~\cite{Sokolov_2021} \\
\bottomrule
\end{tabular}
\label{tab:algorithm_complexity}
\end{sidewaystable}

Depending on the specific ways used to perform purification or dilation, the resulting space and time complexity requirements for quantum simulation will differ. The most space-resource-intensive way is simply to simulate the system and bath altogether as a unitary dynamics on the quantum computer. Examples of this approach are the thermofield double methods~\cite{sagastizabal2021variational,zhu2020generation}, where a system of $n$-qubits will in general need $2n$ qubits. In contrast, the most space-resource-efficient methods will only need one ancillary qubit to block-encode the system Hamiltonian inside a unitary operator, where a polynomial approximation to the partition function can be implemented. In the canonical ensemble for Gibbs state preparation, a polynomial approximation to $e^{-\beta*x}$ is needed (for example, see Ref.~\cite{coopmans2023predicting,powers2023exploring}). This comes at a cost of possibly a long quantum circuit for implementing the block-encoding for generic Hamiltonians. Nevertheless, Ref.~\cite{powers2023exploring} performed such dilation and finite temperature state preparation for a 3-qubit 1D Heisenberg model on real quantum hardware and larger spin systems with a noiseless emulator. A comparison with QITE was also discussed.

Between these two limits, there is often a space-time resource tradeoff in designing proper finite temperature quantum algorithms. We also note the use of a thermal pure quantum state combined with shadow techniques to simplify the initial state preparation in Ref. ~\cite{coopmans2023predicting,powers2023exploring}.
Beyond the circuit width and depth tradeoff, the intrinsic non-unitarity of the partition function also means that most quantum algorithms will suffer from finite success probability, where the cost of amplitude amplification and sampling will need to be considered in evaluating future quantum computing methods.

\begin{figure}[htbp!]
    \centering
    \includegraphics[width=1\linewidth]{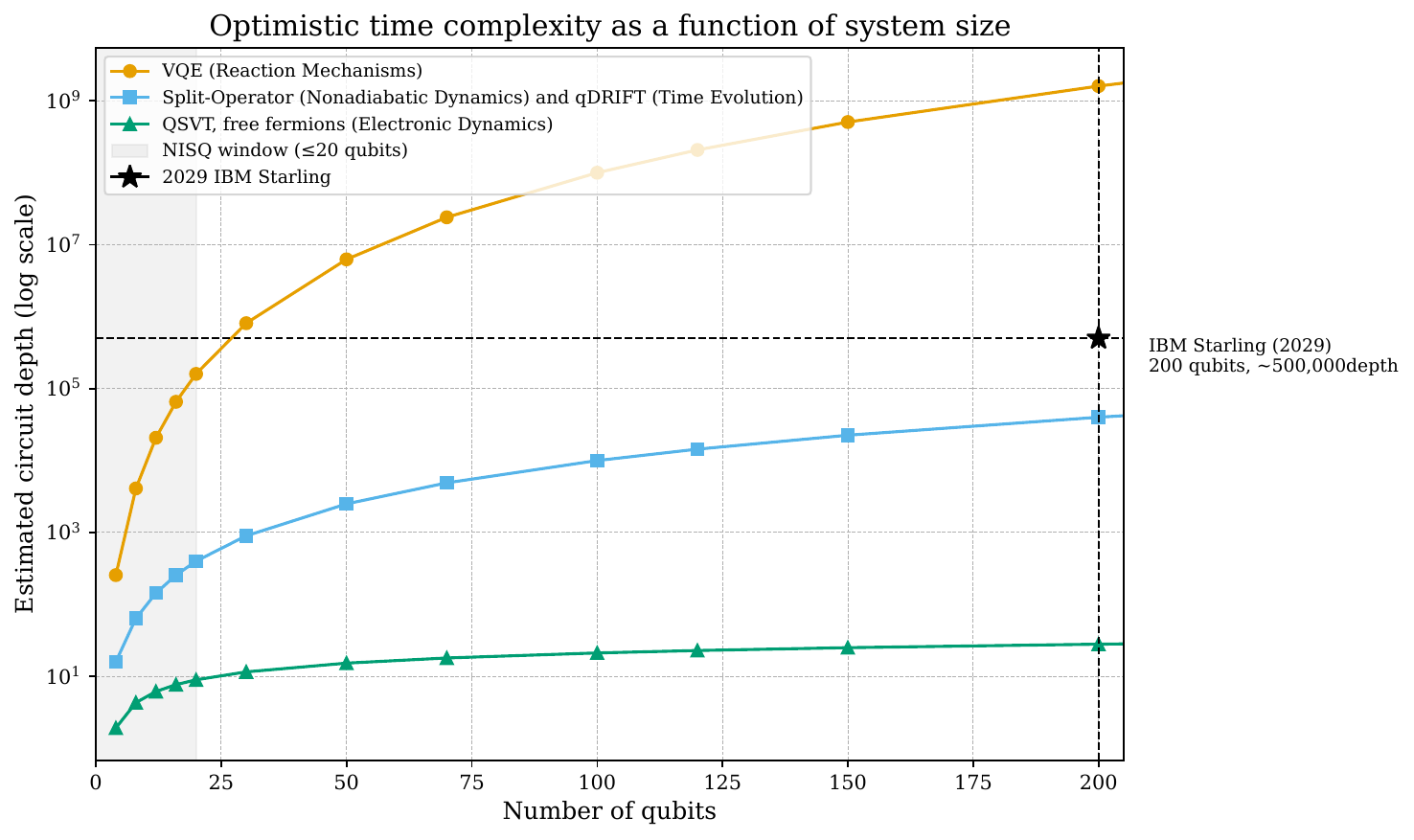}
    \caption{Optimistic logical-depth estimates for representative quantum chemistry algorithms versus system size (number of qubits). 
    Each curve gives the \emph{per-execution} logical circuit depth as a function of the number of logical qubits~$n$, obtained under \emph{best--case} assumptions and with all constant prefactors set to 1.
    Explicit dependences on simulation time~$t$ and target accuracy~$\epsilon$ are suppressed to highlight the qubit–scaling trend.
    For QSVT, we assume a free fermion Hamiltonian with efficient block-encoding, which results in a scaling of $O(log^2(N))$. The grey band marks today’s \emph{NISQ} physical qubit scale ($n \le 20$). The black star and dashed lines indicate IBM’s projected \emph{Starling} processor for 2029, which is projected to have 200 logical qubits with support for $10^8$ gates. $10^8$ gates across 200 qubits implies support for circuits of at least $5\times10^5$ depth circuits \cite{IBM2025Roadmap}.}
    \label{fig:complexity-vs-qubits}
\end{figure}

\section{Conclusions and Future Prospects}
In this manuscript, we have reviewed progress toward modeling quantum chemistry \textit{beyond the ground state} on quantum hardware. A combination of algorithmic and hardware advances have now placed such aims as predicting reaction mechanisms, reactive dynamics, and finite temperature chemistry within reach. Key to enabling these applications are cross-cutting quantum algorithms including the Variational Quantum Eigensolver, which has received significant attention for ground state applications; the Quantum Singular Value Transformation for quantum dynamics, but also has applications in the ground state and other settings; the Linear Combination of Unitaries algorithm; and time evolution algorithms such as qDRIFT and QITE. 

Further progress in enhancing the computational efficiency of these algorithms by reducing the quantum volumes they need or by developing new, more inherently efficient algorithms will accelerate their practical application to problems of chemical significance. A summary of the time complexities of all of many of the algorithms presented in this work may be found in Table~\ref{tab:algorithm_complexity}. We hope this Table sheds light on how different algorithms scale---and potential areas of focus to improve their scaling. As illustrated in Figure~\ref{fig:complexity-vs-qubits}, many of these algorithms can already be employed on small systems using current NISQ hardware. Building on proofs that demonstrate that quantum non-adiabatic and electron dynamics algorithms can confer substantial quantum advantage, we also see from Figure \ref{fig:complexity-vs-qubits} that quantum dynamics algorithms most readily fall within an \textit{optimistic} estimate of the quantum resources that will be available in the next five years. Further algorithmic and hardware advances will be needed to predict reaction mechanisms, particularly in the Born-Oppenheimer approximation, and especially given the sometimes stringent demands of chemical accuracy. Potential avenues for compressing the quantum volumes of related quantum circuits include identifying efficient and accurate active spaces in an automated fashion and/or via chemical intuition, downfolding so that the Hamiltonians only act on qubits of a much smaller active space while implicitly reflecting the influence of the much larger external orbital space~\cite{bauman_downfolding_2019,otten_localized_2022,mitra_localized_2024}, and embedding, in which the system is divided into fragments that can be treated using a high-accuracy method such as quantum computation, while the surroundings are treated using a more efficient classical method~\cite{iijima_towards_2023,ye_bootstrap_2019,liu_bootstrap_2023}. It should be noted that these scaling estimates do not take state preparation, error correction, and measurement costs into full consideration; as discussed in Section~\ref{complexity}, these scalings are non-trivial and may require the greatest innovations to advance the field as whole. 

The aforementioned potential avenues for reducing quantum circuit complexity---and most others---have arisen from years of research into solving quantum chemistry problems on classical computers, but we believe that the greatest advances will arise from completely `quantum' thinking rooted in a deep understanding of not only the non-ground state applications outlined above, but of quantum information. Such thinking remains in its infancy, certainly relative to the decades of research that have gone into the development of quantum chemical algorithms on classical computers. Nonetheless, we look forward to its maturation and the quantum advantage for all of chemistry that we believe it will confer. 

\bibliographystyle{unsrt}
\bibliography{main}

\end{document}